\documentclass[12pt]{iopart}

\usepackage{color}

\expandafter\let\csname equation*\endcsname\relax
\expandafter\let\csname endequation*\endcsname\relax
\usepackage{graphicx}
\usepackage{listings}
\usepackage{subfigure}
\usepackage{color}
\usepackage{float}
\usepackage[T1]{fontenc}
\usepackage{units}
\usepackage{float}
\usepackage{color}
\usepackage{amssymb, amsmath, amsfonts}
\usepackage{amsthm}
\usepackage{xspace}
\usepackage{abraces}
\usepackage{amsmath}

\newcommand*\diff{\mathop{}\!\mathrm{d}\hspace{1pt}}
\newcommand{\dt}{\diff t}

\newcommand{\dx}{\diff x}
\newcommand{\dW}{\diff W}

\newcommand{\dJ}{\diff J}

\begin{document}

\title{Data-Driven Reconstruction of Stochastic Dynamical Equations based on Statistical Moments}

\author{Farnik Nikakhtar$^{1}$, Laya Parkavousi$^{2}$, Muhammad Sahimi$^{3}$,  M. Reza Rahimi Tabar$^{4,5}$, Ulrike Feudel$^{5}$ and Klaus Lehnertz$^{6,7,8}$}
\address{$^{1}$Department of Physics, Yale University, New Haven, CT 06511, USA}
\address{$^{2}$Max Planck Institute for Dynamics and Self-Organization (MPI-DS), 37077, G\"ottingen, Germany}
\address{ $^{3}$Mork Family Department of Chemical Engineering \& Materials Science, University of Southern California, Los Angeles, CA 90089-1211}
\address{$^{4}$Department of Physics, Sharif University of Technology, Tehran 11155-9161, Iran,}
\address{$^{5}$Theoretical Physics/Complex Systems, ICBM, University of Oldenburg, 26129 Oldenburg, Germany}
\address{$^{6}$Department of Epileptology, University of Bonn Medical Centre, Venusberg Campus 1, 53127 Bonn, Germany}
\address{$^{7}$Helmholtz-Institute for Radiation and Nuclear Physics, University of Bonn, Nussallee 14--16, 53115 Bonn, Germany}
\address{$^{8}$Interdisciplinary Center for Complex Systems, University of Bonn, Br\"uhler Stra\ss{}e~7, 53175~Bonn, Germany,}
\ead{klaus.lehnertz@ukbonn.de}
\ead{ulrike.feudel@uni-oldenburg.de}
\vspace{10pt}

\begin{abstract}
Stochastic processes are encountered in many contexts, ranging from generation sizes of bacterial colonies and service times in a queueing system to displacements of Brownian particles and frequency fluctuations in an electrical power grid. 
If such processes are Markov, then their probability distribution is governed by the Kramers-Moyal (KM) equation, a partial differential equation that involves an infinite number of coefficients, which depend on the state variable. 
The KM coefficients must be evaluated based on measured time series for a data-driven reconstruction of the governing equations for the stochastic dynamics. 
We present an accurate method of computing the KM coefficients, which relies on computing the coefficients' conditional moments based on the statistical moments of the time series. 
The method's advantages over state-of-the-art approaches are demonstrated by investigating prototypical one-dimensional stochastic processes with well-known properties.
\end{abstract}


\section{Introduction}
The analysis of time series reflecting the dynamics of complex systems, ranging from physics via engineering to medicine and the neurosciences, needs to be based on assessing the interactions and strengths of random forces operating within the systems. 
Modeling of such systems and time series has been pursued for decades~\cite{ch1bib014, ch1bib015, ch1bib013, ch1013spsAR, tabar2019}. 
Time series with a finite Markov-Einstein scale can be modeled by a data-driven reconstruction of the governing equations, leading to stochastic differential equations (SDE). 
These consist of deterministic and stochastic terms, with the latter corresponding to the fluctuating behavior of the measured time series. 
As described below, the parameters of the SDE, known as Kramers-Moyal (KM) coefficients, are estimated directly from the measured data~\cite{ch1bib013, ch1013spsAR, tabar2019}.

Not only does the modeling aim to understand the physics of the system and to extract as much information from the data as possible, but it also aims to make probabilistic predictions, namely to address the question: given the state of a system at time $t$, what is the probability of finding it in a particular state at time $t+\tau$? 
If the state of a system at time $t$ depends statistically only on its state at the previous time step, but not on earlier ones, then the stochastic process represents a Markov process. 
A given time series may be considered as Markov above the Markov-Einstein scale, which can be estimated directly from observations via some statistical tests~\cite{ch1bib013, ch1013spsAR, tabar2019}. 
 
It is known that the probability distributions --~both marginal and conditional~-- of Markov processes are governed by a first-order partial differential equation in time and an infinite series of derivatives with respect to the state variable $x$. 
The equation is known as the Kramers-Moyal equation~\cite{ch1bib014, ch1bib015} and is given by
\begin{equation}\label{2nd-KM}
\frac{\partial}{\partial t}p(x,t|x_0,t_0)=\displaystyle\sum_{l=0}^\infty\left
(-\frac{\partial}{\partial x}\right)^l\left\{D^{(l)}(x,t)p(x,t|x_0,t_0)\right
\}\;,
\end{equation}
subject to the initial condition $p(x,t_0|x_0,t_0)=\delta(x-x_0)$. 
Here, $D^{(l)}(x,t)$ are the conditional moments of the probability density functions of the transition rates, which --~if all known~-- can identify the probability distribution of the transition rates. 
They are the aforementioned KM coefficients and are given by
\begin{equation} \label{eq:D}
D^{(l)}(x,t)=\lim_{\tau\to 0}\frac{1}{l!}\frac{K^{(l)}(x,t,\tau)}{\tau},
\end{equation} 
with $K^{(l)}(x,t,\tau)$ defined as
\begin{eqnarray} \label{KM}
K^{(l)}(x,t,\tau)&=&\int dx'(x'-x)^l p(x',t+\tau|x,t) \cr \nonumber \\ &=& 
\Big\langle[x(t+\tau)-x(t)]^l\Big|_{x(t)=x}\Big\rangle\;,
\end{eqnarray}
and with $\langle \cdots\rangle$ denoting averaging over the conditional distribution. 
State-of-the-art procedures for estimating the conditional moments $K^{(l)}(x,t,\tau)$ and, henceforth, the KM coefficients involve histogram-based~\cite{tabar2019} and kernel-based regression~\cite{ch17NA,ch17WA,Lam} as well as the maximum likelihood method~\cite{Klein1, Klein2, Garcia}, with the first two methods estimating the conditional probability distribution function (PDF) $p(x',t+\tau|x,t)$ for given $x$ and $x'$ in order to calculate the average in Eq.~(\ref{KM}). 
Other methods are the unbiased estimation method~\cite{Hin} and the statistical inference approach~\cite{kassel}; see also~\cite{Davis} for double kernel-based regression for unevenly sampled time series. 

In general, estimating the conditional PDF using the histogram-based regression with a fixed binning, given the time series with finite number of data points, will lead to a sparse matrix (for $p(x',t+\tau|x,t)$) and, therefore, will not provide reliable results for KM coefficients. 
Furthermore, the kernel-based regression requires an a priori selection of the kernel and its band-width to compute the conditional moments, which represents the method's main obstacle for yielding accurate results. 
In addition to such limitations, due to the sparsity of the matrix of conditional PDFs, such procedures (histogram/kernel) are applicable in one and rarely in two dimensions~\cite{ch1bib013, 2D1, 2D2}. 
On the other hand, similar to other optimization problems, maximum likelihood estimates can be sensitive to the choice of the starting values and are computationally expensive.
To overcome such limitations, we introduce in this paper a new approach, the moments method, for computing the KM coefficients using the short-time correlation functions and statistical moments of time series. 

\section{Estimation of Kramers-Moyal conditional moments using statistical moments} 
\label{sec:HOCM_DP}

If $y_l(t)=[x(t+\tau)-x(t)]^l$ represents the $l$-th increments of $x(t)$, and assuming that the conditional moments in Eq.~(\ref{KM}) can be approximated by an $l'$-order polynomial (see the discussion below for the choice of $l'$), one can write  
\begin{eqnarray}\label{CM1}
&& K^{(l)}(x,t,\tau)=\Big\langle y_l(t,\tau)\Big|x(t)=x\Big\rangle \cr 
\nonumber \\ &&=\phi_0+\phi_1 x+\phi_2 x^2+\phi_3 x^3+\cdots+\phi_{l'}x^{l'},
\end{eqnarray}
where we omitted the dependence of $\phi_i$ on $t$ and $\tau$ to simplify notation. 
We find that the coefficients $\phi_0,\;\phi_1,\cdots,\phi_{l'}$ are the solution of a system of linear equations (see Appendix A)
\begin{equation}\label{set}
\begin{pmatrix}
\langle y_l \rangle \\
\langle xy_l \rangle \\
\vdots \\
\langle x^{l'} y_l \rangle
\end{pmatrix}=
\begin{pmatrix}
1 & \langle x\rangle & \langle x^2 \rangle & \cdots & \langle x^{l'} \rangle \\
\langle x\rangle & \langle x^2\rangle & \langle x^3\rangle & \cdots & \langle 
x^{l'+1}\rangle \\
 \vdots &  \vdots  &  \vdots  &   \vdots & \vdots  \\
\langle x^{l'}\rangle & \langle x^{l'+1} \rangle & \langle x^{l'+2} \rangle & 
\cdots & \langle x^{2l'} \rangle 
\end{pmatrix}
\begin{pmatrix}
\phi_0 \\
\phi_1 \\
\vdots \\
\phi_{l'}
\end{pmatrix},
\end{equation}
where $\langle x^{k} \rangle$ is the statistical moment of $x(t)$ of order $k$.
The left-hand side of Eq.~(\ref{set}) involves terms such as $\langle x^i (x(t+\tau) - x(t))^l \rangle$, which require the estimation of short-time correlation functions of $\langle x^i x(t+\tau)^j \rangle$ and of statistical moments of $\langle x^{i+j} \rangle$. 
Therefore, to compute the KM coefficients, it is necessary to have knowledge about the short-time correlation functions and statistical moments of a time series.
By solving the matrix equation for $\phi_0$, $\phi_1$, $\cdots$, and $\phi_{l'}$, the conditional moments and the KM coefficients of order $l$ are determined by [cf. Eq.~(\ref{eq:D})]
\begin{eqnarray}
D^{(l)}(x,t) & = & \lim_{\tau\to 0}\frac{1}{l!}\frac{1}{\tau}\langle[x(t+\tau)
-x(t)]^l|x(t)=x\rangle \cr \nonumber \\ 
& = & \lim_{\tau\to 0}\frac{1}{l!}\frac{1}{\tau}\left[\phi_0+\phi_1 x+\cdots+
\phi_{l'} x^{l'} \right].
\end{eqnarray}
The $\phi$ coefficients are a function of the time lag $\tau$. 
These coefficients are estimated ($\tilde{\phi}$) for different time lags $\tau = (1,2,3,\ldots) \dt$ and a linear regression is performed for the first four time lags. 
The slopes of the linear regressions are the $\phi$ coefficient.
Here, $\dt$ is the sampling interval of the time series $x(t)$. 
For a non-stationary process, the moments matrix on the right hand side and the vector on the left hand side of Eq.~({\ref{set}}) will be time-dependent, implying that the coefficients $\phi_i$ can also depend explicitly on time (see Appendix B, also for a validation of our method from an analytical perspective).

For a given time series, the order of the polynomial $l'$ is selected based on resolving of $x^{2l'} p(x)$, especially at the tails of $p(x)$. 
By ``resolving'', we mean that for large values of $x$, given a sufficient number of data points,  estimates of $x^{2l'} p(x)$ will be reliable. 
Polynomials of higher orders $l'$ require a larger amount of data points in order to safely estimate the statistical moments $\langle x^{2l'}\rangle$. 
For a given number of data points $n$, increasing $l'$ beyond some upper bound $l'_c$ will no longer resolve $x^{2l'} p(x)$ and, therefore, one should stop the expansion in powers of $x$ 
at order $l'_c$. 
In Appendix C, we determine numerically the upper order of the moments, $l'_c$, for the three examples presented in the following. 

The proposed method is also applicable to any non-polynomial KM coefficients, such as fractional or sinusoidal functions.  
Examples of non-polynomial expressions in Eq.~(\ref{KM}) include the host-immune-tumor model (also known  as dynamical model of cancer growth) and the Kuramoto-Sakaguchi phase oscillator model with higher-order interactions~\cite{new}, to name but a few. 

Before we demonstrate advantages of the proposed method over state-of-the-art approaches, we briefly discuss the truncation of the KM equation~(\ref{2nd-KM}) to rate the number of KM coefficients that should be derived in case of an unknown process. 
The KM equation offers three possibilities: (i) truncating the expansion at $l=1$ represents a deterministic process; (ii) truncating at $l=2$ results in the Fokker-Planck equation for diffusion processes; and (iii) including all terms up to $l \to \infty$. 
Checking the condition $D^{(l)}(x,t) = 0$ for $l \geq 3$ in a given time series is not a straightforward task as it requires knowledge of the KM conditional moments $K^{(l)}(x, t,\tau)$ in powers of the sampling time $\tau$. 
Lehnertz {\it et al}.  have shown that for a process to be classified as diffusive, the condition $K^{(4)}(x, t,\tau) \simeq 3 (K^{(2)}(x, t, \tau))^2$ must be satisfied~\cite{njp}. 
If this holds, then the first two KM coefficients are sufficient to describe the measured stochasticity of the time series. 
Otherwise, the dynamics can be approximated using a jump-diffusion process which requires knowledge of the KM coefficients of orders 1, 2, 4, and 6~\cite{ch12bib012,tabar2019}.

In what follows, we reconstruct one-dimensional stochastic dynamical equations from three prototypical stochastic processes with well-known properties, of which two examples represent diffusion processes and the third one is a jump-diffusion process. 

We generated synthetic time series by a numerical simulation of the corresponding dynamical equations using the adaptive integration method for It\^o SDEs~\cite{ch14SDE8,ch14SDE9}. 
Then, we estimated the KM coefficients $D^{(l)}(x,t)$ of the time series using four methods, namely histogram-based and kernel-based regressions, maximum likelihood estimation, and the moments methods proposed in this paper. 
For each example, the standard error of the mean of KM coefficients was computed based on $100$ realizations of the applied noises and with $n=10^6$ data points. 
The mean-squared error (MSE) of the estimated as well as the coefficient of determination ($R^2$ score) between estimated and actual values for the drift and diffusion coefficients
(the first two KM coefficients) and the jump parameters, i.e., amplitude and jump-rate, are reported in Table~\ref{table:data}. \\

\begin{figure}
\center
{\includegraphics[width=\textwidth]{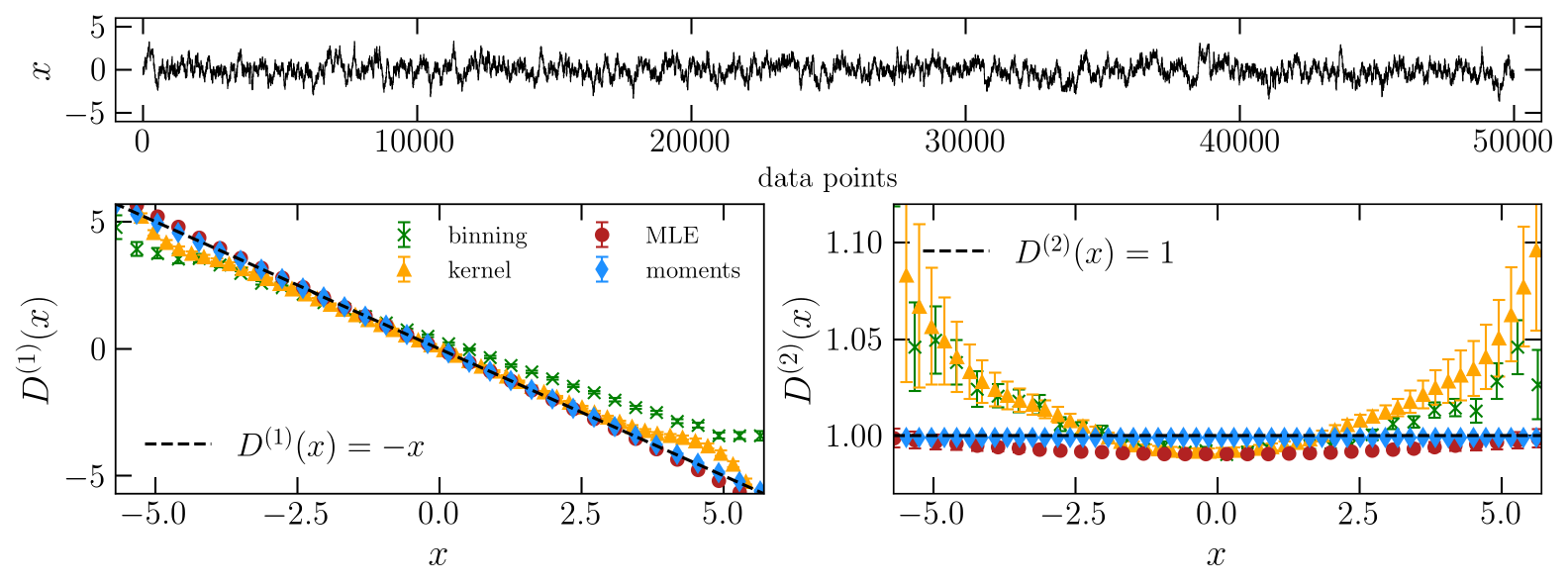}}
\caption{Upper panel: Excerpt of an exemplary time series of an 
Ornstein--Uhlenbeck process (Eq.~(\ref{langevin}) with drift and diffusion coefficients $D^{(1)}(x)=
-x$ and $D^{(2)}(x)=1$. Lower panels: Theoretical (dashed lines) and  
estimated drift coefficients using histogram-based and kernel-based 
regressions, maximum likelihood estimation (MLE), and the proposed moments method. We 
used the best-tuned values (minimum MSE and maximum $R^2$ score) for the number of bins and band-width for histogram-based and the kernel-based regressions, 
which we determined to be $M=45$ and $h\simeq 0.3$ (with a Gaussian kernel). 
We chose the order of polynomials for estimating the drift and diffusion coefficients to be $l'=3$ and $l'=4$, respectively (cf. Appendix C). 
Error bars are the standard error of the mean, estimated from 100 realizations of synthetic time series.}
\label{Ex1}
\end{figure}

\noindent
{\it Example 1: Diffusion processes: Ornstein--Uhlenbeck process.} 
The Pawula theorem~\cite{pawula1967} states that the KM expansion can be truncated at  $l=2$ if $D^{(4)}({\bf x},t)$ vanishes. 
The KM equation then reduces to the Fokker-Planck equation, with the state variable $x(t)$ describing the diffusion process. 
In that case, $x(t)$ will be governed by the Langevin equation which (using the It\^o description) has the following form:
\begin{equation}\label{langevin}
\dx=D^{(1)}({x},t) \dt +\sqrt{2D^{(2)}({x},t)}\dW(t)\;,
\end{equation}
where $D^{(1)}({x},t)$ and $D^{(2)}({x},t)$ are the drift and diffusion coefficients, respectively. 
$\{W(t),t\geq 0\}$ is a scalar Wiener process. 

We consider the Langevin equation~(\ref{langevin}) with $D^{(1)}(x)=-x$, $D^{(2)}(x)=1$, and integrate it with $\dt=0.01$ to generate synthetic time series. 
Our findings for $D^{(1)}(x)$ and $D^{(2)}(x)$ are shown in Fig.~(\ref{Ex1}). 
For the linear process considered, the MSEs of $D^{(1)}(x)$ and $D^{(2)}(x)$ estimated with the  moments method are smaller (with $R^2$ closer to unity) than the respective values estimated with the other methods (Table~\ref{table:data}). 
For an analytical solution of Eq.~({\ref{set}}) for the coefficients $\phi_i$ and the KM coefficients, as well as for the condition to stop increasing the order of moments, see Appendices B and C.\\

\begin{figure}
{\includegraphics[width=\textwidth]{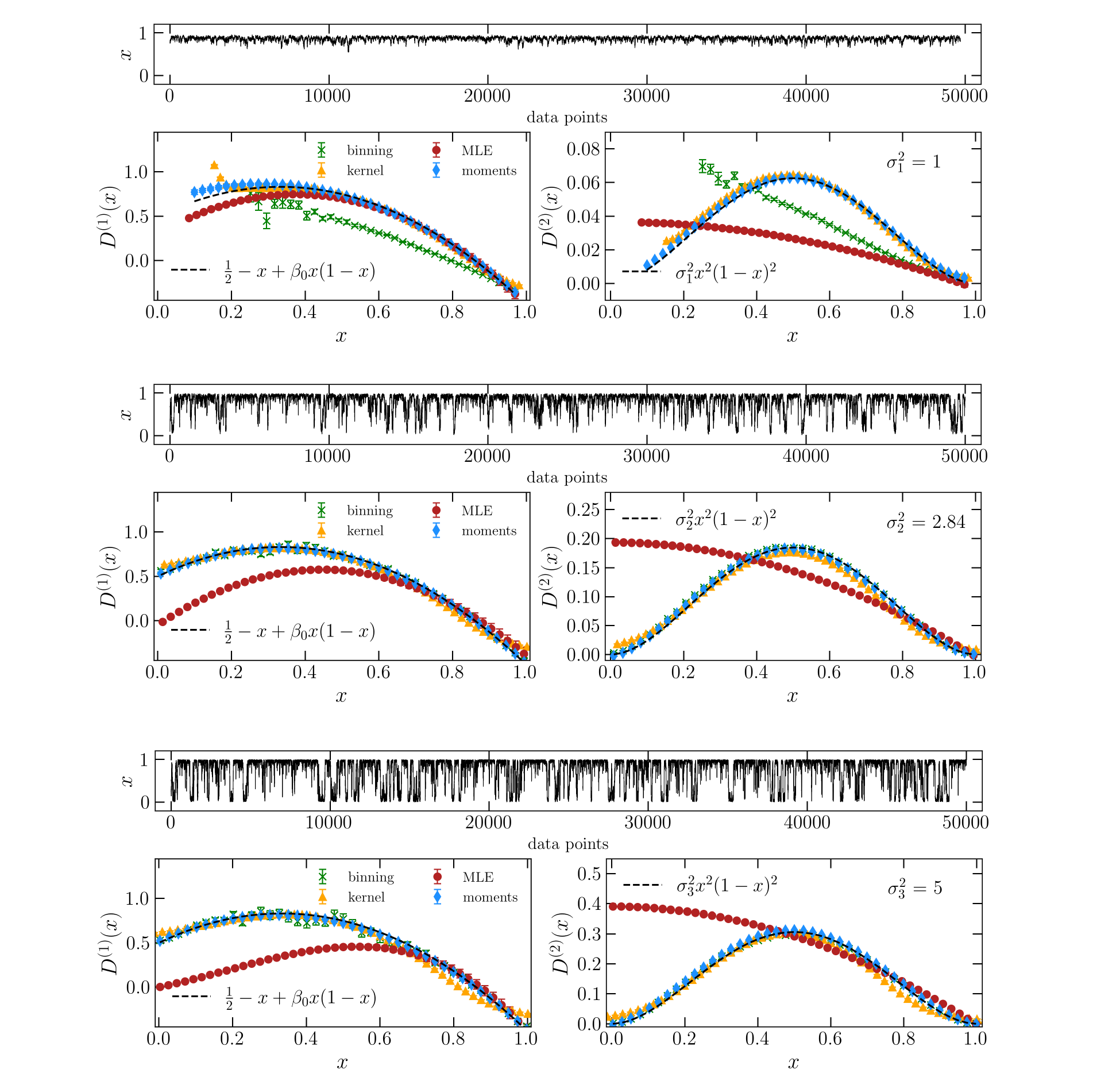}}
\caption{Top to bottom: Excerpts of exemplary time series of the noisy genetic model (Eq.~(\ref{genic22})) as well as drift and diffusion coefficients $D^{(1)}(x)=\frac{1}{2}-x+\beta_0 
x(1-x)$ (with $\beta_0=3$) and $D^{(2)}(x)=\sigma^2 x^2(1-x)^2$ for $\sigma^2<2.84$ (top), $\sigma^2 \approx 2.84$ (middle), and $\sigma^2>2.84$ (bottom). 
Theoretical (dashed lines) and the estimated drift and diffusion coefficients using the four methods. 
The best-tuned values for the number of bins and for the kernel band-width were $M=41$ and $h\simeq0.3$ (with a Gaussian kernel). 
The order of polynomials for estimating the drift and diffusion coefficients were $l'=3$ and $l'=4$, respectively (cf. Appendix C). 
Error bars are the standard error of the mean (SEM) estimated from 100 realizations of the synthetic time series.}
\label{Ex2}
\end{figure}

\begin{table*}[b]
\centering
\scalebox{0.6}{
\begin{tabular}{ | c | c || c | c | c || c | c | }
\hline
example & drift coefficient MSE/${R^2}$ & diffusion coefficient MSE/${R^2}$ & jump rate MSE/${R^2}$ & jump 
amplitude MSE/${R^2}$ & method  \\ 
\hline \hline
Ornstein-Uhlenbeck process & 0.2168/0.7133 & 0.0001/-17.2624 & N/A & N/A & histogram-based \\ 
\hline
Ornstein-Uhlenbeck process & 0.0106/0.9915 & 0.0001/-0.0623 & N/A & N/A & 
kernel-based
\\ \hline
Ornstein-Uhlenbeck process & 0.0001/0.9999 & 0.0001/-9.2731 & N/A & N/A & 
MLE
\\ \hline
Ornstein-Uhlenbeck process &  0.0001/0.9999 & 0.0001/0.9999 & N/A 
& N/A  & moments \\ 
\hline \hline
noisy genetic model ($\sigma < \sigma_c$) & 0.0567/0.4069 & 0.0003/0.3845 & N/A & N/A &  histogram-based \\ 
\hline
noisy genetic model ($\sigma<\sigma_c$)  & 0.0041/0.9708 & 0.0001/0.9739 & N/A & N/A & kernel-based\\ 
\hline
noisy genetic model ($\sigma<\sigma_c$)  & 0.0055/0.9474 & 0.0005/-3.3394 & N/A & N/A & MLE \\ 
\hline
noisy genetic model ($\sigma<\sigma_c$) & 0.0011/0.9921 & 0.0001/0.9865 & N/A & N/A  & moments\\ 
\hline \hline
noisy genetic model ($\sigma\simeq\sigma_c$) & 0.0007/0.9949 & 0.0001/0.9931 & 
N/A & N/A &  histogram-based \\ 
\hline
noisy genetic model ($\sigma \simeq \sigma_c$) & 0.0029/0.9784 & 0.0001/0.9712 & N/A & N/A & kernel-based\\ 
\hline
noisy genetic model ($\sigma \simeq \sigma_c$) & 0.0684/-0.0221 & 0.0058/-0.6734 & N/A & N/A & MLE\\ 
\hline
noisy genetic model ($\sigma \simeq \sigma_c$) &  0.0001/0.9999 & 0.0001/0.9972 & N/A & N/A & moments\\ 
\hline \hline
noisy genetic model ($\sigma > \sigma_c$) & 0.0015/0.9891 & 0.0001/0.9962 & N/A & N/A &  histogram-based\\ 
\hline
noisy genetic model ($\sigma > \sigma_c$) & 0.0091/0.9429 & 0.0003/0.9739 & N/A & N/A & kernel-based\\
\hline
noisy genetic model ($\sigma > \sigma_c$) & 0.1322/-1.3872 & 0.0263/-0.8569 & N/A & N/A & MLE\\
\hline
noisy genetic model ($\sigma > \sigma_c$) &  0.0003/0.9977 & 0.0001/0.9959 & N/A & N/A & moments\\ 
\hline \hline
jump-diffusion process
& 0.1684/-2.3667 & 0.0003/-0.0035 & 0.0199/-0.2823 & 0.0013/-0.3256 &  
histogram-based\\ 
\hline
jump-diffusion process 
& 0.0782/0.4234 &  0.0013/-1.0006  & 0.0032/-1.1983 & 0.0002/-0.3163
& kernel-based\\ 
\hline
jump-diffusion process 
& --/-- &  --/--  & --/-- & --/-- 
& MLE\\ 
\hline
jump-diffusion process
&  0.0001/0.9999 & 0.0001/-0.5863 & 
0.0057/-0.3062 & 0.0002/-0.3062  & moments\\ 
\hline \hline
\end{tabular}
}
\caption{Mean squared errors (MSEs) of estimated and $R^2$ scores between estimated and actual values of the drift and diffusion coefficients, as well as jump rate and jump amplitude for the examples considered.
Results are given for the histogram-based and the kernel-based regressions, maximum likelihood estimation (MLE), and the proposed moments method. 
Since no theoretical expression for the short-time propagator of $p(x',t+\tau|x,t)$ is known for the jump-diffusion processes, MLE could not be applied.
MSEs and $R^2$ scores for the Ornstein--Uhlenbeck process were estimated in the interval $x\in [-2,2]$, for the noisy genetic model in the interval $x\in [0.1,1]$, and for the jump-diffusion process 
in the interval $x\in [-1.5, 1.5]$.}
\label{table:data}
\end{table*}

\noindent
{\it Example 2: Diffusion processes: the noisy genetic model.} 
As a second example from the family of diffusion processes, we consider a nonlinear multiplicative process, namely, the noisy genetic model in the context of population genetics~\cite{Wio}, which exhibits a noise-induced transition from unimodal to bimodal probability distribution functions with increasing noise intensity. 
The noisy genetic model is described as a one-variable diffusive SDE~\cite{Arnold} (see Appendix D)
\begin{equation}\label{genic22}
\dx=\left[\alpha-x+\beta_0 x(1-x)\right]\dt+\sigma x(1-x)\dW(t)\;, 
\end{equation}
where $x\in [0,1]$, and $\alpha\in(0,1)$ stands for the mutation rate. 
Here, $\beta_0$ denotes the gene selection factor and, independent of $\alpha$ and $\beta_0$, the deterministic part of the dynamical system $(\ref{genic22})$ has only one equilibrium point in the corresponding interval, which is stable. 
Moreover, the intensity $\sigma$ is a real constant. 

There is a critical $\sigma^2_c$ such that for $\sigma^2<\sigma^2_c$, the stationary solution of the corresponding Fokker-Planck equation, $p_s(x)$, has only one extremum, namely a maximum. 
When $\sigma^2>\sigma^2_c$, the stationary probability distribution has three extrema, with $x_2$ being a minimum and $x_{1}$ and $x_{3}$ each being a maximum. 
Therefore, when $\sigma^2<\sigma^2_c$, $p_s(x)$ is unimodal at $\sigma^2=\sigma^2_c$, the probability distribution becomes flat, followed by a transition to bimodal distribution functions for more intense noise. 

We integrate Eq.~(\ref{genic22}) with $\dt=0.01$ to generate synthetic time series for $\sigma_1^2<\sigma^2_c$, $\sigma_3^2>\sigma^2_c$, and $\sigma_2^2 \approx \sigma^2_c$. Our results for the drift and diffusion coefficients are summarized in Fig.~(\ref{Ex2}).
We find that for all values of $\sigma^2$, the MSEs corresponding to the proposed moments method are comparably smaller (with $R^2$ closer to unity), as presented in Table~\ref{table:data}.\\

\noindent
{\it Example 3: Jump-diffusion process with a bistable potential.} 
It has become evident that a non-vanishing $D^{(4)}({\bf x},t)$ is a signature of jump discontinuities in a time series~\cite{ch12bib012,njp}. 
In this case, one needs the KM coefficients of at least order six, $D^{(6)}({\bf x},t)$, and in some other cases, up to order eight~\cite{tabar2019}) to estimate jump amplitude and rate, and the KM equation will contain all the terms up to $l\to\infty$~\cite{ch12bib5sps0,ch12bib5sps1,ch12bib5sps2,ch12bib5sps3, ch12bib012, ch12bib5sps4,ch12bib5sps5,ch12bib5sps6,ch12bib5sps7,ch12bib5sps8,
ch12bib5sps9}. 
For non-vanishing $D^{(4)}({\bf x},t)$, one models a time series as a jump-diffusion process, which is given by a (It\^o) dynamical stochastic 
equation~\cite{tabar2019, ch12bib012}
\begin{equation}\label{JDD2}
\dx(t)=D^{(1)}(x,t)\dt+\sqrt{D(x,t)}\dW(t)+\xi \dJ(t)\;,
\end{equation}
where $\{W(t),t\geq 0\}$ is a scalar Wiener process, $D^{(1)}(x,t)$ is the drift coefficient, $D(x,t)$ is the diffusion function, and $J(t)$ is a Poisson jump process. 
The jump rate is $\lambda(x)$ that can be state-dependent with size $\xi$, which we assume to have some symmetric distribution with finite even-order statistical moments $\langle \xi^{2j} \rangle$ for $j\geq 1$. 
For a jump-diffusion process~(\ref{JDD2}), all functions and parameters can be determined based on measured time series by estimating the KM conditional moments~\cite{ch12bib012, tabar2019} (see Appendix E). 
For simplicity, we assume that the jump size $\xi$ has a Gaussian PDF with variance $\sigma_\xi^2$. 
In that case, $\langle\xi^{2k+1}\rangle=0$ and $\langle\xi^{2k}\rangle=\frac{2n!}{2^kk!}\langle\xi^2\rangle^k$.

\begin{figure*}
{\includegraphics[width=\textwidth]{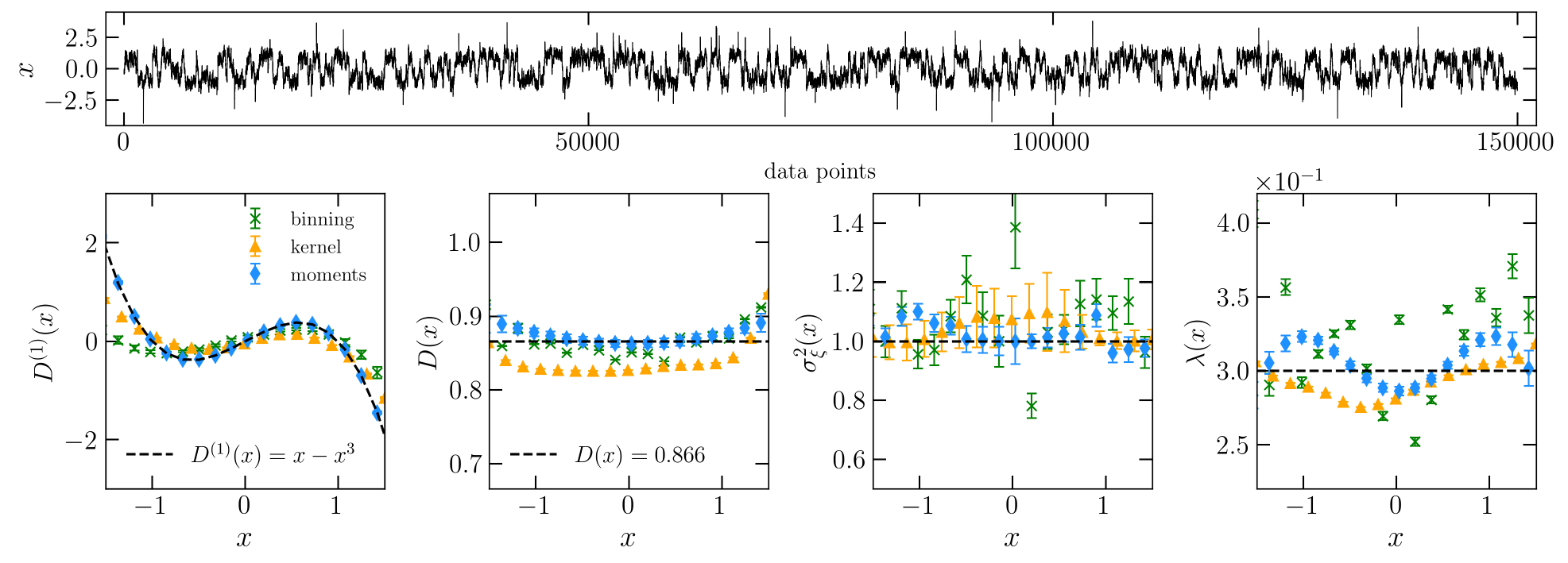}}
\caption{Top: Excerpt of an exemplary time series of a jump-diffusion process (Eq.~(\ref{JDD2})) with drift and diffusion coefficients $D^{(1)}(x)=x-x^3$ and $D(x)=\sqrt{0.75}\approx 0.866$,
and with jump amplitude $\sigma_{\xi}^2=1$ and jump rate $\lambda=0.3$. 
Theoretical (dashed lines) and estimated functions and parameters of the jump-diffusion model using the four methods are shown in the lower panels. 
The best-tuned values for the number of bins and the kernel band-width were $M=61$ and $h\simeq0.3$ (with a Gaussian kernel).
We expanded the KM coefficients $D^{(4)}(x,t)$ and $D^{(6)}(x,t)$ with up to the fourth- and sixth-order polynomials ($l'=4$ and $l'=6$), respectively (cf. Appendix C). 
Error bars are the standard error of the mean, estimated from 100 realizations of the synthetic time series. 
Note, that we do not report results for MLE here and in Table~\ref{table:data},  since a theoretical expression for the short-time propagator $p(x',t+\tau|x,t)$ --~which is required for MLE~-- is not known for the jump-diffusion processes, in contrast to the general Langevin equation~\cite{Klein2}.}
\label{Ex3}
\end{figure*}

We consider a dynamical system with a cubic nonlinearity ($D^{(1)}(x,t) = x-x^3$) with additive white noise, $D(x,t)=\sqrt{0.75}$, jump amplitude $\sigma_\xi^2=1$, and jump rate $\lambda=0.3$. 
The deterministic part of the dynamics has three fixed points at $0$, one unstable fixed point, and $\pm 1$ as stable ones. 
In Fig.~(\ref{Ex3}), we show the numerical values for both the reconstructed drift and diffusion coefficients, as well as jump amplitude and jump rate. 
We find that the MSEs corresponding to the proposed moments method are comparably smaller, see Table~\ref{table:data}. 

\section{Conclusions}  
\label{seq:concl}

We introduced a new method to estimate conditional averages of increments $(x(t+\tau)-x(t))^l$, with $l=0,1,\cdots$, that are needed for a data-driven reconstruction of stochastic dynamical equations from experimental time series. 
The method consists of determining a conditional average in terms of statistical moments of the measured time series. 
It replaces the estimation of the sparse conditional probability distributions with solving a system of linear equations that has polynomial time complexity.
Using three paradigmatic linear and nonlinear stochastic processes, we demonstrated that mean-squared errors (as well as coefficients of determination of KM coefficients) estimated with the proposed method are smaller (larger) than the respective values estimated with current state-of-the-art methods.

While the examples presented in this work focus on stationary moments, we acknowledge the potential extension of our theoretical approaches to non-stationary processes (as discussed in Appendix B). 
We note that our method has not been implemented for non-stationary data at this stage. 
However, we are actively working on proving the method's applicability to time-dependent Fokker-Planck equations and expanding its scope to non-stationary time series. 
We intend to provide further updates and findings in the near future.
As a final remark, we note that the proposed method is easily extendable to multivariate time series, which will enable one to characterize pairwise, as well as higher-order interactions in 
the dynamics~\cite{new}. 
Our moments method provides a way to rapidly and reliably estimate functions and parameters needed for a reconstruction of dynamic equations of complex systems.

\newpage 
\appendix
\noindent

\setcounter{section}{0}

\section{Proof of relation (5)}
\label{ap:A}

For two random variables $y(t)$ and $x(t)$ with joint probability density function $p(y,x)$, the conditional expectation of $y(t)$, given that $x(t)=x$, is given by 
\begin{equation}
\langle y(t)|x(t)=x\rangle=\int yp(y|x)dy\,
\end{equation}
where $p(y|x)$ is the conditional probability, which is obtained via $p(x,y)$ and $p(x)$ using Bayes' theorem. 
In particular, by setting $y_l(t)$ as $(x(t+\tau)-x(t))^l$, the conditional Kramers-Moyal (KM) moments will be given by 
\begin{equation}\label{CM}
K^{(l)}(x,t,\tau)=\Big\langle[x(t+\tau)-x(t)]^l\Big|x(t)=x\Big\rangle,
\end{equation}
in accordance with Eq. (3) in the main text~\cite{ch1bib014}. 
Let us assume that the conditional moments $K^{(l)}(x,t,\tau)$ have a functional form $A_l(x)$, whose parameters depend implicitly on $l$ and the time-lag $\tau$ and time $t$. 
By definition of conditional expectation and using Bayes' theorem, we write
\begin{equation}\label{CE-cub-main1}
\int y p(y|x)dy = \int y\frac{p(y,x)}{p(x)}dy=  A_l(x)
\end{equation}
for a given function $A_l(x)$, implying that 
\begin{equation}\label{CE-cub-main2}
\int y {p(y,x)}dy=  p(x)  \cdot A_l(x).
\end{equation}
By integrating both sides of Eq.~(\ref{CE-cub-main2}) over $x$, one finds,
\begin{equation}\label{CE-cub-main3}
\int dx\int y{p(y,x)}dy=\int dy yp(y)=\int dxp(x)\cdot A_l(x).
\end{equation}
$A_l(x)$ can have any functional form~\cite{new}. 
An extreme case for the time series is white noise, for which $K^{(l)}(x,t,\tau)$ will be a polynomial of order $l$. 

Let us consider $A_l(x)$ as a polynomial function of $x$ of order $l'\geq l$. 
Therefore, we must determine $l'+1$ free parameters, and we 
have 
\begin{equation}
\langle y(t)|x(t)=x\rangle=\phi_0+\phi_1 x+\phi_2 x^2+\phi_3 x^3+\cdots+ 
\phi_{l'} x^{l'},
\end{equation}
where we can choose $y_l \equiv y_l(t,\tau)=(x(t+\tau)-x(t))^l$ for a given $l$. 
Using Eq.~(\ref{CE-cub-main2}), one finds
\begin{eqnarray}\label{CE-cub-main}
& & \int y_lp(x,y_l)dy=\cr\nonumber \\
& & \phi_0 p(x)+\phi_1 xp(x)+\cdots+\phi_{l'} x^{l'} p(x). 
\end{eqnarray}
To determine the unknown coefficients $\phi_i$, $i=0,\ldots,l'$ in Eq.~(\ref{CE-cub-main}), we need $l'+1$ independent relations. For the first of our systems of equations, we integrate both sides of the above equation over variable $x$ to obtain
\begin{equation}
\langle y_l\rangle=\phi_0+\phi_1\langle x\rangle+\cdots+\phi_l'\langle x^{l'} 
\rangle.
\end{equation}
Multiplying Eq.~(\ref{CE-cub-main}) by $x$, $x^2$, $\cdots$, $x^{l'}$ and integrating over $x$, we obtain a system of linear equations
\begin{equation}\label{set2}
\begin{pmatrix}
\langle y_l \rangle \\
\langle xy_l \rangle \\
\vdots \\
\langle x^{l'} y_l \rangle
\end{pmatrix}=
\begin{pmatrix}
1 & \langle x\rangle & \langle x^2 \rangle & \cdots & \langle x^{l'} \rangle \\
\langle x\rangle & \langle x^2\rangle & \langle x^3\rangle & \cdots & \langle 
x^{l'+1}\rangle \\
 \vdots &  \vdots  &  \vdots  &   \vdots & \vdots  \\
\langle x^{l'}\rangle & \langle x^{l'+1} \rangle & \langle x^{l'+2} \rangle & 
\cdots & \langle x^{2l'} \rangle 
\end{pmatrix}
\begin{pmatrix}
\phi_0 \\
\phi_1 \\
\phi_2 \\
\vdots \\
\phi_{l'}
\end{pmatrix}
\end{equation}
which proves the relation~(5) in the main text.

\newpage
\section{Analytical solutions}
\label{ap:B}

\noindent
{{\it Example 1: Stationary process, Ornstein--Uhlenbeck process.}}\\

As we mentioned in the main text in the first example, the Ornstein--Uhlenbeck (OU) process can be written as a Langevin equation of the form
\begin{equation}\label{OU}
    \dx(t)= - \theta x(t) \dt + \mu \dW(t)\;,
\end{equation}
with drift and diffusion coefficients given by  $D^{(1)}(x,t) = -\theta x $, and $\sqrt{2D^{(2)}(x,t)} = \mu$. Here $\mu$ and $\theta$ are positive constants. 
For $\theta>0$, the process~(\ref{OU}) has a stationary probability distribution~\cite{tabar2019}. The mean of the process can be set to zero and the covariance of $x(t)$ and $x(s)$ in the stationary state can be written as~\cite{tabar2019}
\begin{equation}
    \text{cov}(x(t), x(s)) = \frac{\mu^2}{2\theta} \exp{(-\theta |t - s|)}.
\end{equation}
The statistical moments of the OU process would be as follows
\begin{equation}
    \langle x^k \rangle = \begin{cases} (k-1)!! \left(\frac{\mu^2}{2\theta}\right)^{k / 2} & k \text { even } \\ 0 & k \text { odd }\end{cases}.
\end{equation}
The double factorial is the product of all the integers from 1 up to $k$ that have the same parity as $k$. 

To demonstrate our moments method, let us assume that the conditional KM moments are a cubic function of $x$, so by having the analytical form of the covariance and statistical moments, the matrix elements of the system of the linear equations (Eq.~\ref{set2}) can be determined as
\begin{equation}\label{set3}
\begin{pmatrix}
\langle y \rangle \\
\langle xy \rangle \\
\langle x^{2} y \rangle \\
\langle x^{3} y \rangle
\end{pmatrix}=
\begin{pmatrix}
1 & 0 & \frac{\mu^2}{2\theta} & 0 \\
0 & \frac{\mu^2}{2\theta} & 0 & 3(\frac{\mu^2}{2\theta})^2 \\
\frac{\mu^2}{2\theta} & 0 & 3(\frac{\mu^2}{2\theta})^2 & 0  \\
0 & 3(\frac{\mu^2}{2\theta})^2 & 0 & 15(\frac{\mu^2}{2\theta})^3
\end{pmatrix}
\begin{pmatrix}
\phi_0 \\
\phi_1 \\
\phi_2 \\
\phi_3
\end{pmatrix}.
\end{equation}
For the first order KM moment (to obtain the drift coefficient), we set $y \equiv y_1(t,\tau) = x({t+\tau}) - x(t)$, so the vector in the l.h.s. of Eq.~(\ref{set3}) would be obtained as
\begin{equation}\label{set4}
\begin{pmatrix}
0 \\
\frac{\mu^2}{2\theta} (e^{-\theta \tau} - 1) \\
0\\
3 (\frac{\mu^2}{2\theta})^2 (e^{-\theta \tau} - 1)
\end{pmatrix}.
\end{equation}
The solution for the parameters vector in this system of linear equations is
\begin{equation}\label{sol1}
\begin{pmatrix}
\phi_0 \\
\phi_1 \\
\phi_2 \\
\phi_3
\end{pmatrix} =
\begin{pmatrix}
0 \\
e^{-\theta \tau} - 1 \\
0\\
0
\end{pmatrix},
\end{equation}
which means $K^{(1)} = (e^{-\theta \tau} - 1) x(t)$, and the drift coefficient is
\begin{eqnarray}
    D^{(1)}(x,t) &=& \lim _{\tau \rightarrow 0} \frac{K^{(1)}(x,t)}{\tau}  \\ \cr \nonumber  &=& \lim _{\tau \rightarrow 0} \frac{1}{\tau} (1 - \theta \tau -1 +  \mathcal{O}(\tau^2)) ) x(t) = -\theta x(t).
\end{eqnarray}

For the second KM moment (to obtain the diffusion coefficient), we set $y\equiv y_2(t,\tau) =  (x({t+\tau}) - x(t))^2$, so the vector in the l.h.s. of Eq.~(\ref{set3}) and the solution to the linear system would be as follows
\begin{equation}\label{set4}
\begin{pmatrix}
\frac{\mu^2}{\theta} (1 - e^{-\theta \tau}) \\
0 \\
\frac{\mu^2}{2\theta} (4 - 3e^{-\theta \tau} + 2e^{-2\theta \tau}) \\
0
\end{pmatrix},
\end{equation}
and
\begin{equation}\label{sol2}
\begin{pmatrix}
\phi_0 \\
\phi_1 \\
\phi_2 \\
\phi_3
\end{pmatrix} =
\begin{pmatrix}
\frac{\mu^2}{\theta} (1 - e^{-\theta \tau}) \\
0 \\
0 \\
0
\end{pmatrix}.
\end{equation}
Note that the moments matrix in the r.h.s. of Eq.~(\ref{set3}) is not a function of $y$ and does not change.
Therefore, the diffusion coefficient is
\begin{eqnarray}
    D^{(2)}(x,t) &=& \frac{1}{2} \lim _{\tau \rightarrow 0} \frac{K^{(2)}(x,t)}{\tau}
    \\ \cr \nonumber  &=& 
        \lim _{\tau \rightarrow 0} \frac{1}{2\tau} \frac{\mu^2}{\theta} (1 - 1 + \theta \tau +  \mathcal{O}(\tau^2))) = \frac{\mu^2}{2}.
\end{eqnarray}

\vskip 0.7cm

 \noindent
{{\it Example 2: Non-stationary process, Wiener process.}}\\

As an example of a non-stationary process, we consider the Wiener process whose derivative is a white noise. It is known that the Wiener process has a vanishing drift and a constant diffusion coefficient which equals to $1/2$~\cite{tabar2019}.  
The covariance of a Wiener process is
\begin{equation}
    \text{cov}(x(t), x(s)) = \min{(t, s)}\;,
\end{equation}
where $\min(t, s)$ is the minimum of two chosen times $(t, s)$.
With $x(0)=0$ and $t_0=0$ the statistical moments of Wiener process are given by
$\langle x^{2k+1}\rangle = 0$ and
\begin{equation}
\langle x^{2k}\rangle =  \frac{\Gamma(k+1/2)} {\Gamma(1/2)} ~  2^k t^{k}  \hskip 0.5cm k=1,2,\ldots \;, \nonumber
\end{equation}
where $\Gamma(x)$ is the Euler gamma function. {Here $\langle \cdots\rangle$ denotes ensemble averaging.}

For the moments matrix we have
\begin{equation}
\begin{pmatrix}
1 & \langle x\rangle & \langle x^2 \rangle & \langle x^{3} \rangle \\
\langle x\rangle & \langle x^2\rangle & \langle x^3 \rangle & \langle x^{4} \rangle \\
\langle x^2\rangle & \langle x^3\rangle & \langle x^4 \rangle & \langle x^{5} \rangle  \\
\langle x^3\rangle & \langle x^4\rangle & \langle x^5 \rangle & \langle x^{6} \rangle
\end{pmatrix} = 
\begin{pmatrix}
1 & 0 & t & 0 \\
0 & t & 0 & 3t^2 \\
t & 0 & 3t^2 & 0  \\
0 & 3t^2 & 0 & 15t^3
\end{pmatrix}.
\end{equation}

To find the drift coefficient, we have to solve the following linear set of equations with $y\equiv y_1(t,\tau)=(x(t+\tau)- x(t))$:
\begin{equation}\label{sol3}
\begin{pmatrix}
\langle y \rangle \\
\langle xy \rangle \\
\langle x^{2} y \rangle \\
\langle x^{3} y \rangle
\end{pmatrix}=
\begin{pmatrix}
0 \\
0 \\
0 \\
0
\end{pmatrix} = 
\begin{pmatrix}
1 & 0 & t & 0 \\
0 & t & 0 & 3t^2 \\
t & 0 & 3t^2 & 0  \\
0 & 3t^2 & 0 & 15t^3
\end{pmatrix}
\begin{pmatrix}
\phi_0 \\
\phi_1 \\
\phi_2 \\
\phi_3
\end{pmatrix},
\end{equation}
This gives us $(\phi_0, \phi_1, \phi_2, \phi_3) = (0, 0, 0, 0)$, and then $D^{(1)}(x,t)$ will be
\begin{equation}
    D^{(1)}(x,t) = \lim _{\tau \rightarrow 0} \frac{K^{(1)}(x,t)}{\tau} = 0.
\end{equation}

Similarly, to find the diffusion coefficient one has to solve the following set of linear equations with
$y\equiv y_2(t,\tau)=(x(t+\tau)- x(t))^2$:
\begin{equation}\label{sol3}
\begin{pmatrix}
\langle y \rangle \\
\langle xy \rangle \\
\langle x^{2} y \rangle \\
\langle x^{3} y \rangle
\end{pmatrix}=
\begin{pmatrix}
\tau \\
0 \\
t \tau \\
0
\end{pmatrix}=
\begin{pmatrix}
1 & 0 & t & 0 \\
0 & t & 0 & 3t^2 \\
t & 0 & 3t^2 & 0  \\
0 & 3t^2 & 0 & 15t^3
\end{pmatrix}
\begin{pmatrix}
\phi_0 \\
\phi_1 \\
\phi_2 \\
\phi_3
\end{pmatrix},
\end{equation}
This gives $(\phi_0, \phi_1, \phi_2, \phi_3) = (\tau, 0, 0, 0)$. 
To compute higher-order covariances in the l.h.s. of Eq.~(\ref{sol3}), we have used Wick's theorem~\cite{tabar2019}
\begin{eqnarray}
\langle x(i)^3 x(j) \rangle & =&3 \langle x(i)^2 \rangle \langle x(i) x(j) \rangle \\
\langle x(i)^2 x(j)^2 \rangle & =& \langle x(i) x(i) \rangle  \langle x(j) x(j) \rangle + 2 \langle x(i) x(j) \rangle^2.
\end{eqnarray}
Finally, the diffusion coefficient is given by
\begin{equation}
    D^{(2)}(x,t) = \frac{1}{2} \lim _{\tau \rightarrow 0} \frac{K^{(2)}(x,t)}{\tau} = \frac{1}{2} ~.
\end{equation}

\newpage
\section{Stop-condition for the order of polynomials in the expansion of the Kramers-Moyal coefficients}
\label{ap:C}

\begin{figure*}
{\includegraphics[width=\textwidth]{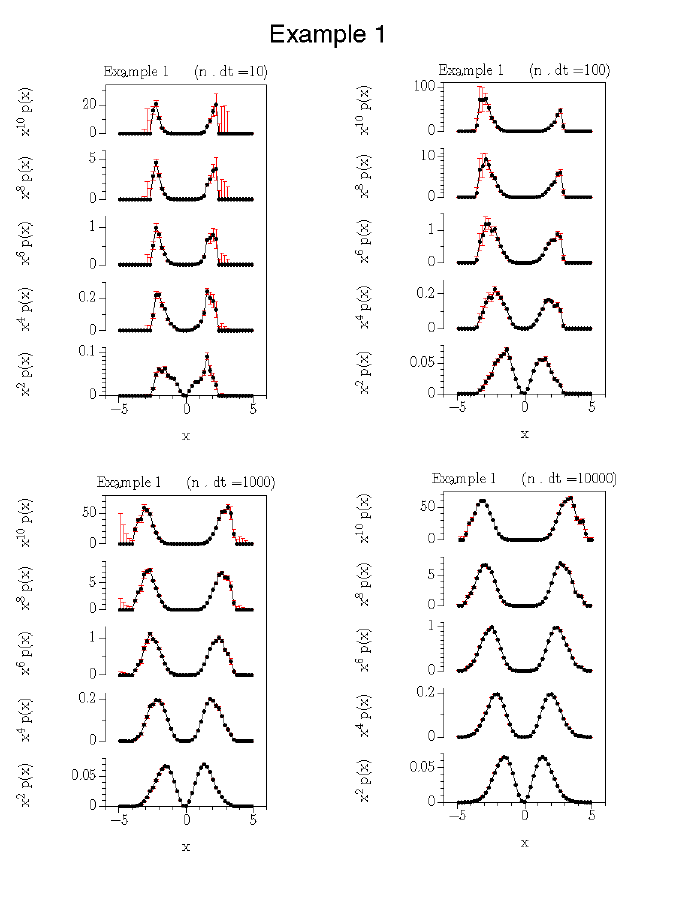}}
\caption{Plots of $x^2~ p(x)$, $\cdots$, $x^{10}~ p(x)$, estimated from time series of example 1 for different $n \dt$. We estimate the error bars from 10 realizations of the generated time series.}
\label{PDF1}
\end{figure*}

\begin{figure*}
{\includegraphics[width=\textwidth]{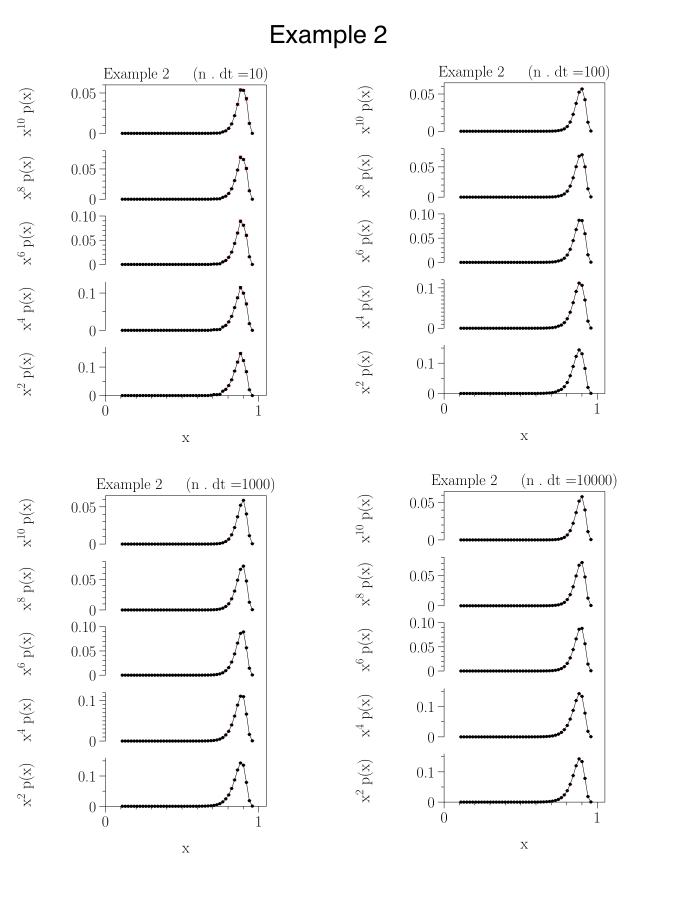}}
\caption{Same as Fig. (\ref{PDF1}) but for example 2.}
\label{PDF2}
\end{figure*}

\begin{figure*}
{\includegraphics[width=\textwidth]{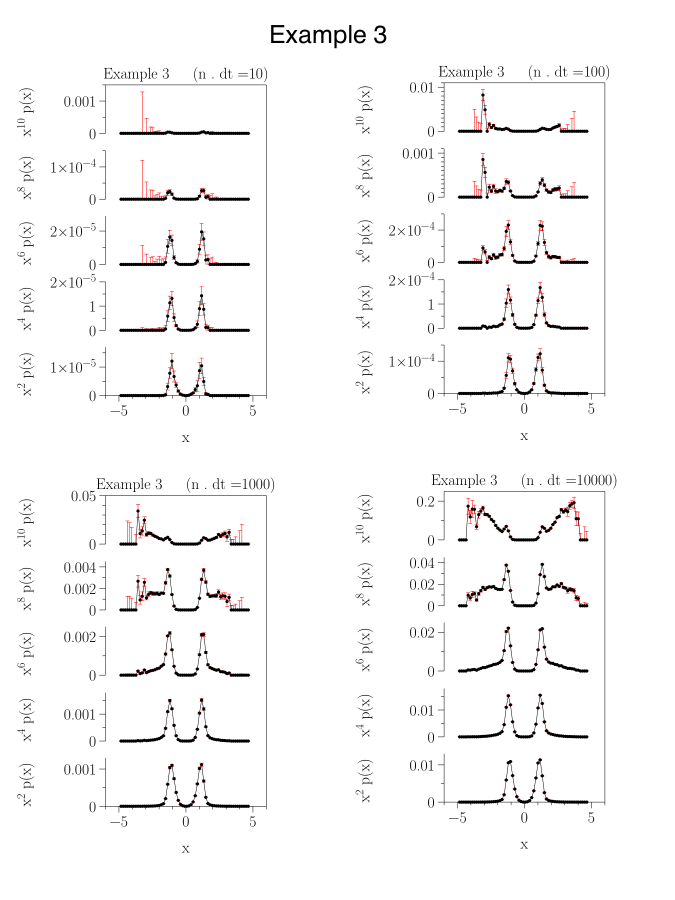}}
\caption{Same as Fig. (\ref{PDF1}) but for example 3.}
\label{PDF3}
\end{figure*}

As shown in relation~(\ref{set2}), to estimate the KM coefficients with a polynomial of order $l'$ one needs to estimate statistical moments of $x$ up to the order $2 l'$.
To study how such moments can be reliably estimated from given time series (each with length $n \dt$), one should ensure that the tails of $ x^l ~ p(x)$ (for $l=1,2,\cdots, 2l'$) are well resolved.
In what follows, we show how one can check the reliability of the estimation of the tail of the PDF for the three examples considered in the main text, where the drift and diffusion coefficients are polynomials of orders 1, 2, and 3, respectively. For these examples, we need to estimate the statistical moments at least up to the order of six.  
In Figs.~(\ref{PDF1}), ~(\ref{PDF2}), and~(\ref{PDF3}), we plot $x^2~ p(x)$, $\cdots$, $x^{10}~ p(x)$, estimated from time series of the three examples and for different $n \dt$. 
For given $n \dt \in \left\{10, 100, 1000, 10000\right\}$, we estimate the error bars from 10 realizations of the generated time series for each example. 
As shown in Fig.~(\ref{PDF1}), for example 1 and with increasing $n \dt$ the tails of PDF of $p(x)$ are well-resolved, rare events are detected, and $x^{2k}~ p(x)$ possesses smaller error bars (in the size of symbols), particularly in the tails.  
Similar discussions are valid for the other examples. 
We note that in the main text, the results have been reported from ensemble averaging over 100 realizations of time series for each example.

\begin{figure*}
{\includegraphics[width=\textwidth]{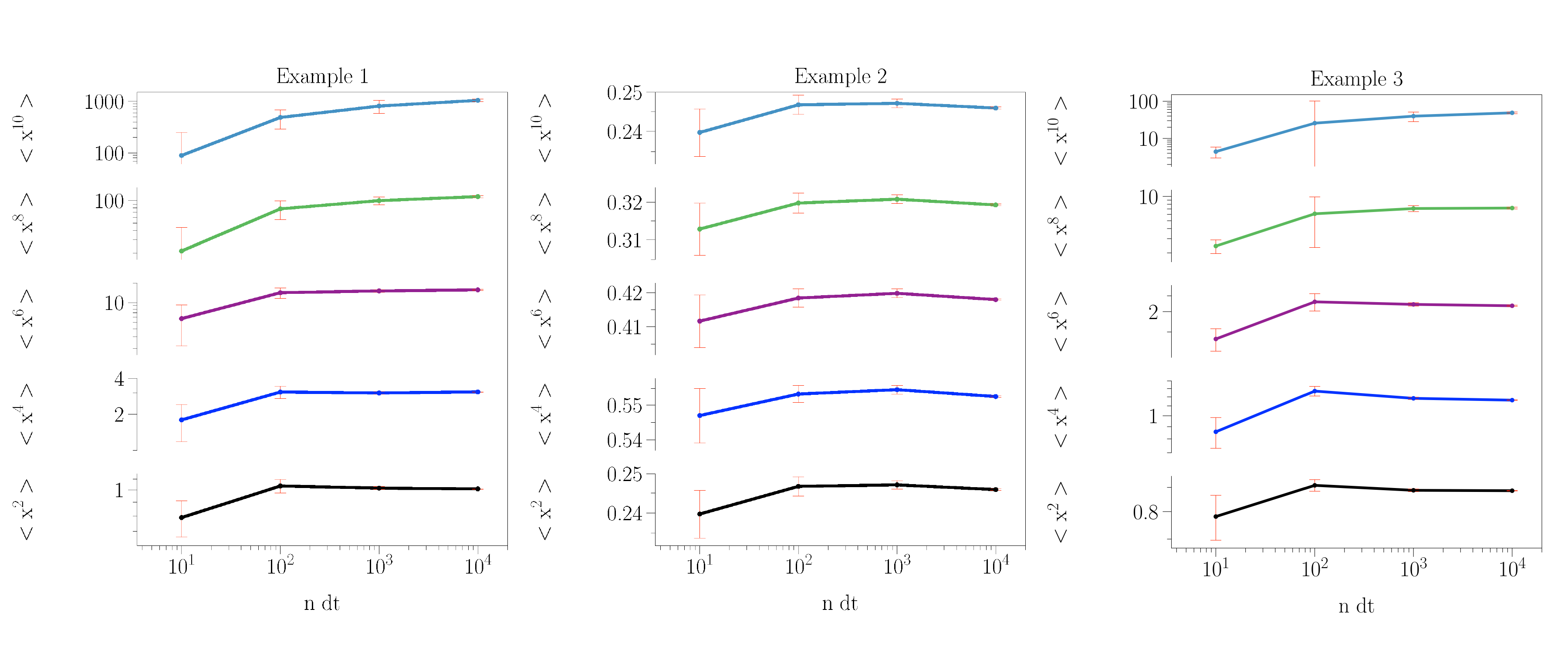}}
\caption{Dependence of  statistical moments $\langle x^{2k} 
\rangle$, with $k=1,2,\cdots,5$ on $n\dt$ from 10 realizations for various $n\dt$ for the three examples of the main text. The middle panel is for example 2 with $\sigma_1^2<\sigma^2_c$.}
\label{M}
\end{figure*}

\begin{figure*}
{\includegraphics[width=\textwidth]{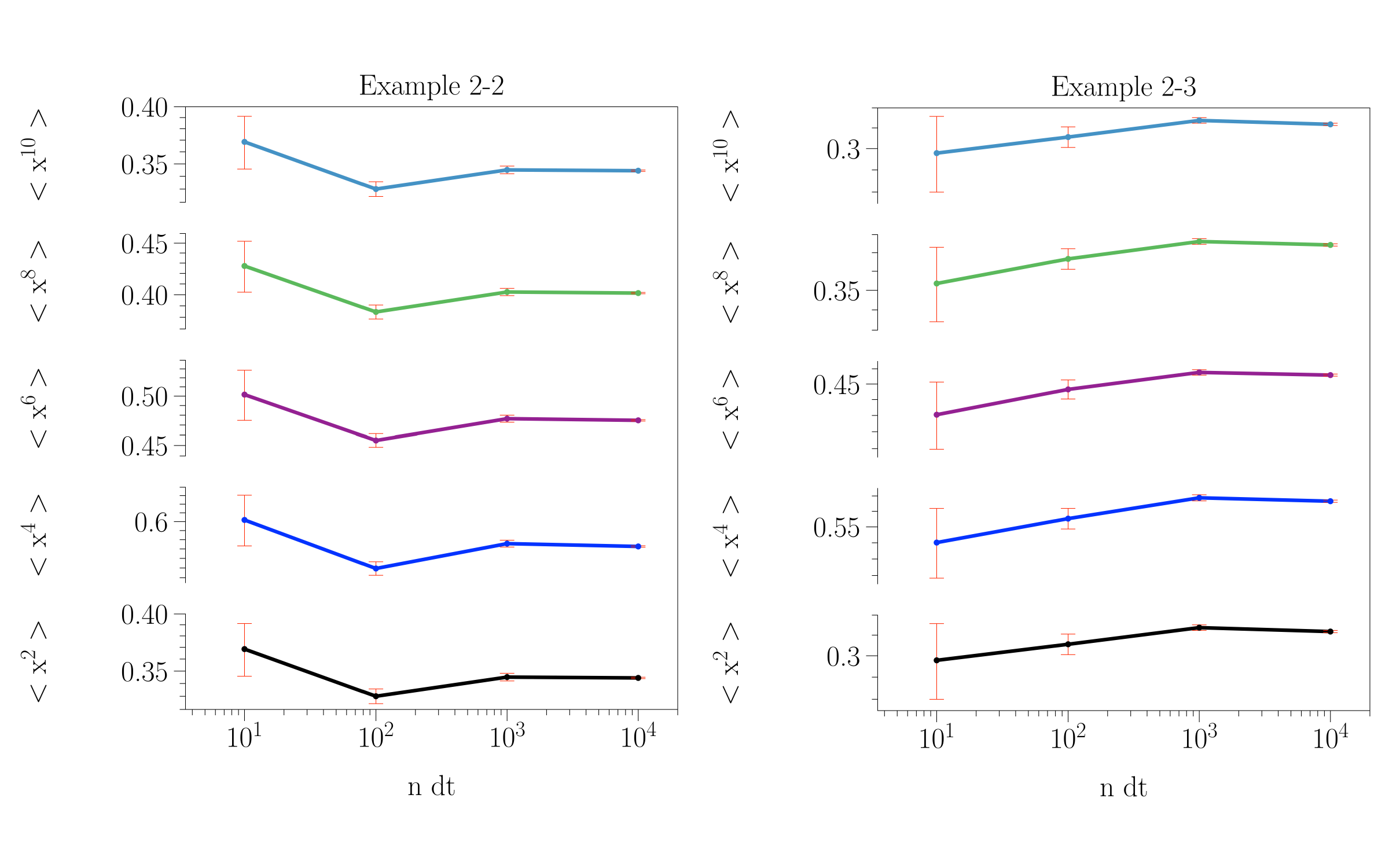}}
\caption{ Same as Fig. (\ref{M}) but for example 2 with $\sigma_2^2 \approx \sigma^2_c$ (left) and with $\sigma_3^2>\sigma^2_c$ (right).}
\label{ME}
\end{figure*}

In addition, in Figs.~(\ref{M}) and~(\ref{ME}), we plot the statistical moments $\langle x^{2k} \rangle$, with $k=1,2,\cdots,5$ (from 10 realizations) for different $n \dt$. 
Considering example 1, for $n\dt>1000$ the statistical moment  $\langle x^{8} \rangle$ and those of higher orders are not approaching constant values up to $n\dt=10000$. 
Therefore by considering the error bars one can choose $l'=6$. 

For a given stationary time series, we can analyze the statistical moments $\langle x^{2k} \rangle$ with varying numbers of data points $n$. 
In this case, the error bars are calculated based on a number (e.g., 100) of randomly selected data chunks, each consisting of $n$ data points from the time series.

\begin{figure*}
{\includegraphics[width=\textwidth]{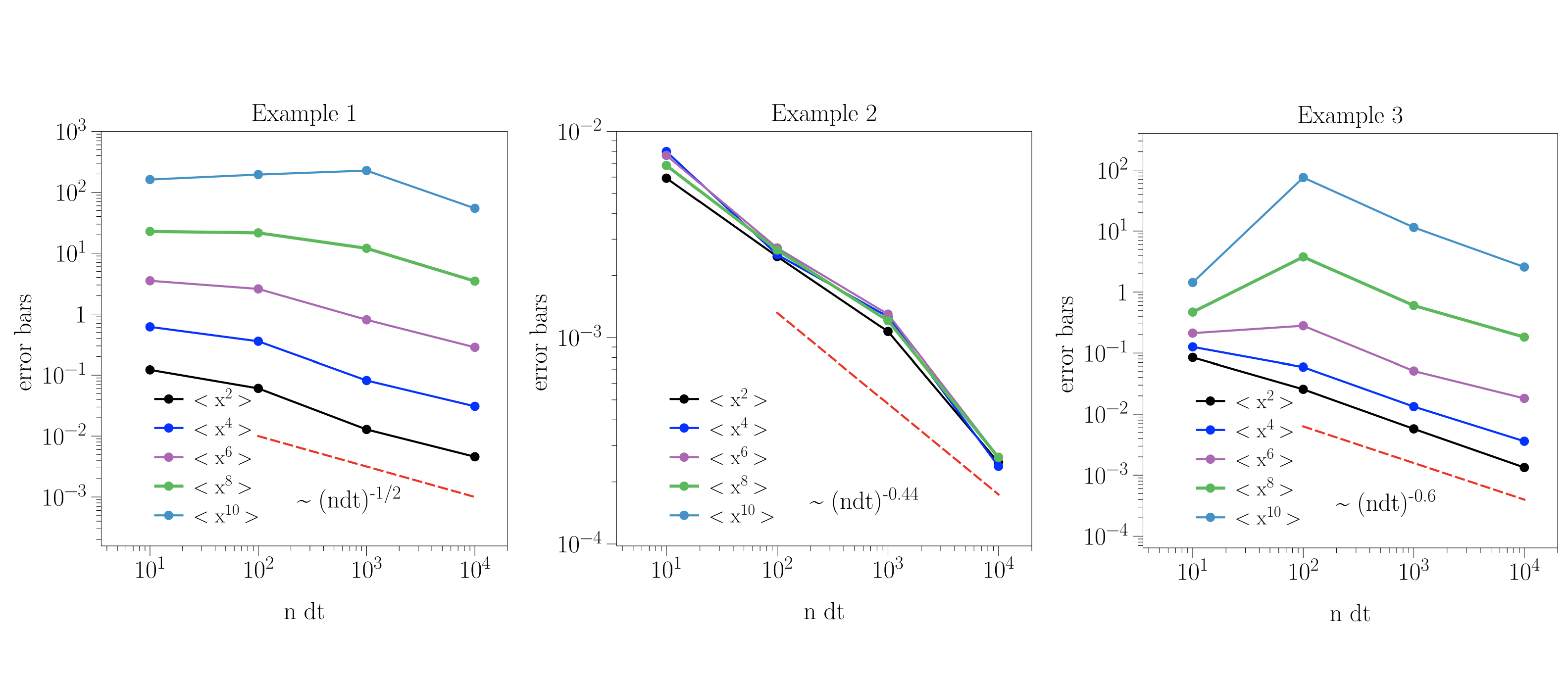}}
\caption{Errors of estimations of $\langle x^{2k} \rangle$, with $k=1,2,\cdots,5$ (from 10 realizations) for different $n \dt$, for the three examples. Errors decrease with $n \dt$ as (red dashed line) $ {1}/{(n \dt)^\gamma}$, with $\gamma\simeq 0.5$.}
\label{E}
\end{figure*}

In summary, using the saturating behavior of  $\langle x^{2k} \rangle$ for different $k$ with increasing $n \dt$, one can decide in a data-driven way that the KM coefficients can be approximated with a polynomial of order $l'_c$.
In Fig.~(\ref{E}), we plot the errors of estimations of $\langle x^{2k} \rangle$, with $k=1,2,\cdots,5$ (from 10 realizations) for different $n \dt$ for the three examples. 
The errors decrease with $n \dt$ as $ {1}/{(n \dt)^\gamma}$, with $\gamma\simeq 0.5$. 

Finally, we note that in cases where we have access to ground truth, we can provide specific details about the minimum order of the polynomial regression. 
To determine the appropriate minimum order, it is important to utilize a distance measure that allows us to select the minimum order without sacrificing essential information required for accurate estimation of the KM coefficients. 
By reducing the polynomial order, e.g. from 4 to 3, and subsequently to 2 and 1, we can assess the impact on the accuracy of coefficient estimation.
The choice of the distance measure depends on the specific context and analysis requirements. 
The ultimate objective is to identify the minimum polynomial order that preserves the necessary information for accurate estimation of the KM coefficients, while avoiding unnecessary complexity.

\clearpage
\newpage
\section{The noisy genetic model}
\label{ap:D}

We consider a haploid group $x$ as our data and suppose that each haploid may have genes $A$ or $B$. 
The simple genotype selection model can be described as a one-variable deterministic differential equation~\cite{Arnold}
\begin{equation} \label{genic}
\frac{dx}{dt} = \alpha - x + \beta x (1-x), 
\end{equation} 
where $x$ is the ratio between the number of genes $A$ and the total number of genes (A+B), so that $x\in [0,1]$, and $\alpha\in(0,1)$ stands for the mutation rate of gene B. 
Here, $\beta$ denotes the gene selection factor, where $\beta=1$ means that the selection is completely propitious to gene $A$ haploid, whereas $\beta=-1$ implies that the selection is completely propitious to gene $B$ haploid. 
Independent of $\alpha$ and $\beta$, the dynamical system $(\ref{genic})$ has only one equilibrium point in the corresponding interval given by 
\begin{equation}\label{genic1}
x_s=\frac{\beta-1+\sqrt{(\beta-1)^2+4\alpha\beta}}{2\beta},
\end{equation} 
which is stable. 
Now, consider the system~(\ref{genic}) coupled to a noisy environment, with $\beta$ fluctuating. 
We assume that $\beta(t)=\beta_0+\sigma \eta$, where $\eta$ is a white noise with unit intensity and $\sigma$ is a real constant. 
The Langevin equation (noisy genetic model) is given by
\begin{equation}\label{genic2}
\dx=\left[\alpha-x+\beta_0 x(1-x)\right]\dt+\sigma x(1-x)\dW(t)\;, 
\end{equation}
where $\dW(t)=\eta(t)\dt$ and $W(t)$ is a Wiener process. 
The noisy genetic model has two stationary points, that can be obtained by solving the corresponding Fokker-Planck (FP) equation, one of which coincides with the stationary point determined from the 
deterministic component (the drift coefficient) of the dynamics. 
The stationary solution of the FP equation of the Langevin dynamics~(\ref{genic2}) for $\alpha=1/2$ is
\begin{equation}\label{fp}
p_s(x)=\frac{C}{x(1-x)}\exp\left[-\frac{1}{\sigma^2}\left\{\frac{1}{2x(1-x)}+ 
\beta_0\ln\left(\frac{1-x}{x}\right)\right\}\right]\;,
\end{equation}
where $C$ is a normalization constant. 

There is a critical $\sigma^2_c$ in Eq.~(\ref{fp}) such that for $\sigma^2<\sigma^2_c$, the stationary solution of the FP equation, $p_s(x)$, is unimodal, i.e., it has only one extremum, namely a maximum. 
At $\sigma^2=\sigma^2_c$, the probability distribution becomes flat, followed by a transition to a bimodal probability distribution function for more intense noise. 
In fact, the model expressed by Eq.~(\ref{genic2}) exhibits a noise-induced transition with changing $\sigma$. 
The critical $\sigma_c^2$ depends on $\beta_0$. 
In order to have a non-trivial drift (non-linear) in~(\ref{genic2}), we consider the case $\beta_0\neq 0$, say $\beta_0=3$. 
The critical noise intensity in this case is $\sigma^2_c \simeq 2.84$. For $\sigma^2<\sigma^2_c$ (we choose $\sigma^2=1$), $x\approx 0.885$ is the only extremum, which is a maximum. 
For $\sigma^2>\sigma^2_c$ (we choose $\sigma^2=5$), the extrema are located at $x_1\approx 0.023$, $x_2\approx 0.435$, and $x_3\approx 0.982$, with $x_{1}$ and $x_{3}$ each being a maximum and $x_2$ is the minimum.

\newpage
\section{Functions and Parameters of the Jump-Diffusion Process}
\label{ap:E}

For the jump-diffusion process, which is described by a dynamical stochastic equation~(9) (in the main text), all functions and parameters
can be determined based on measured time series by estimating the KM conditional moments~\cite{ch12bib012, tabar2019}:
\begin{eqnarray} \label{eq:KMcoeffspsest11sps12}
K^{(1)}(x,t) & = & D^{(1)}(x,t)\tau \cr \nonumber \\
K^{(2)}(x,t) & = & [D(x,t)+\lambda(x,t)\langle\xi^2\rangle]\tau\cr \nonumber \\
K^ {(j)}(x,t)& = & \lambda(x,t)\langle\xi^j\rangle\tau,\;\; {\rm for}\;\; j>2,
\end{eqnarray}
where the KM conditional moments are $K^{(l)}(x,t,\tau)=\langle[x(t+\tau)-x(t)]^l|_{x(t)=x}\rangle$, as given in Eq.~(3) in the main text. 

Using the last relation in Eq.~(\ref{eq:KMcoeffspsest11sps12}) with $j=4$ and $j=6$, we first estimate the 
state-dependent (Gaussian) jump amplitude $\sigma_{\xi}^2(x)$ and jump rate $\lambda(x)$ via
\begin{eqnarray}\label{eq:jar}
\sigma_{\xi}^2(x,t)=\frac{M^{(6)}(x,t)}{5 M^{(4)}(x,t)}, {\hskip
0.5cm} \lambda(x,t)=\frac{M^{(4)}(x,t)}{3\sigma_{\xi}^4(x,t)},
\end{eqnarray}
where 
$M^{(l)}(x,t)=\lim_{\tau\to 0}\frac{K^{(l)}(x,t,\tau)}{\tau}$. We remind that $D^{(l)}(x,t)$ is related to $M^{(l)}(x,t)$ as $ D^{(l)}(x,t) = M^{(l)}(x,t)/l!$. 

Once the jump components $\sigma_{\xi}^2 (x,t)$ and $\lambda (x,t)$ are identified, the second moment $M^{(2)}(x,t)=D(x,t)+\lambda(x,t)\langle\xi^2 \rangle$ identifies the diffusion function $D(x,t)$, while the first moment yields the estimate for the drift function $D^{(1)}(x,t)$; see Eq.~(\ref{eq:KMcoeffspsest11sps12}).
In practice, we can also define the averaged jump amplitude and jump rate as $\overline{\sigma^2 _\xi} = \frac{1}{N_r} \sum_{k=1} ^{N_r} \sigma^2 _\xi(x_k)$ and $\overline{\lambda} = \frac{1}{N_r} \sum_{k=1} ^{N_r} \lambda(x_k)$, where $N_r$ is the number of data points in some range of the state variable over which the coefficients are being evaluated. 
For stationary processes, we estimate the diffusion coefficient $D(x)$ from the relation $M^{(2)}(x) \simeq D(x)  + \overline{\sigma^2 _\xi} ~ \overline{\lambda}$~\cite{tabar2019}. 
We note that the jump rate per unit of time will be $\lambda_{\text{unit time}}(x)= \lambda(x) \dt$. \\

\newpage
\section*{References}

\bibliographystyle{iopart-num}
\providecommand{\newblock}{}

\end{document}